\documentclass[prl,twocolumn,showpacs,preprintnumbers,amsmath,amssymb]{revtex4}

\usepackage{graphicx}
\usepackage{ifthen}
\usepackage{ifpdf}
\usepackage{color}

\usepackage{latexsym}
\usepackage{amssymb}
\usepackage{bm}
\newcommand{\half}{\mbox{\small $\frac{1}{2}$}}

\newcommand{\eexp}{\mbox{e}^}
\newcommand{\bra}{\left\langle}
\newcommand{\ket}{\right\rangle}

\newcommand{\beq}[1]{\begin{eqnarray}\ifthenelse{#1=-1}{\nonumber}
{\ifthenelse{#1=0}{}{\label{e#1}}}}
\newcommand{\eeq}{\end{eqnarray}}
\newcommand{\hide}[1]{}

\begin{document}

\title{Rings and boxes in dissipative environments}
\author{Yoav Etzioni{$^1$}, Baruch Horovitz{$^1$} and Pierre Le Doussal{$^2$} }

{\affiliation{{$^1$} Department of Physics, Ben Gurion University,
Beer Sheva 84105 Israel}
 \affiliation{{$^2$} CNRS-Laboratoire de
Physique Th{\'e}orique de l'Ecole Normale Sup{\'e}rieure, 24 rue
Lhomond,75231 Cedex 05, Paris France.}

\begin{abstract}
We study a particle on a ring in presence of a dissipative Caldeira-Leggett environment and derive its response to a DC field. We find, through a 2-loop renormalization group analysis, that a large dissipation parameter $\eta$ flows to a fixed point $\eta_R=\eta_c=\hbar/2\pi$. We also reexamine the mapping of this problem to that of the Coulomb box and show that the relaxation resistance, of recent interest, is quantized for large $\eta$. For finite $\eta>\eta_c$ we find that a certain average of the relaxation resistance is quantized. We propose a box experiment to measure a quantized noise.

\end{abstract}


\maketitle
Two of the most important mesoscopic structures are rings, for the study of persistent currents, and quantum dots or boxes, for the study of charge quantization. Of particular recent interest is the quantization of the relaxation resistance $R_q$, defined via an AC capacitance of a single electron box (SEB). Following the prediction of B\"{u}ttiker, Thomas and Pr\^{e}tre \cite{buttiker} that $R_q=h/2e^2$ for a single mode resistor, a quantum mesoscopic RC circuit has been implemented in a two-dimensional electron gas \cite{gabelli} and $R_q=h/2e^2$ has been measured. The theory has been recently extended to include Coulomb blockade effects \cite{mora} showing that $R_q=h/2e^2$ is valid for small dots and crosses over to $R_q=h/e^2$ for large dots.

In parallel, recent data has observed Aharonov-Bohm oscillations from single electron states in semiconducting rings \cite{kleemans}. Further theoretical works have considered the effects of dissipative environments on a single particle in a ring \cite{guinea}, in particular studying the renormalization of the mass $M^*$ and its possible relation to dephasing \cite{guinea,golubev1,kagalovsky,bh}.

It is rather remarkable that the ring and box problems are related via the AES mapping \cite{AES} where the ring experiences a Caldeira-Leggett (CL) \cite{CL} environment. While the exact mapping assumes weak tunneling into the box with many channels, it has been extensively used to describe various tunnel junctions \cite{schon}, the Coulomb blockade phenomena in SEB and in the single electron transistor (SET) \cite{schon,golub,hofstetter,beloborodov,herrero,bulgadayev,lukyanov1,lukyanov2,burmistrov1,burmistrov2,burmistrov3}.

In the present work we address the ring problem by the real time Keldysh method and study it using a 2-loop expansion and renormalization group (RG) reasoning. We find that perturbation theory identifies an unexpected new small parameter $\sin(\frac{\hbar}{2\eta})$ where $\eta$ is the dissipation parameter on the ring, or the lead-dot coupling in the SEB. We infer that a large $\eta$ flows to a fixed point $\eta_R=\eta_c$ with $\hbar/2\eta_c=\pi$. An intuitive argument for this quantization is given before the conclusions. For large $\eta$ our RG is consistent with 2 loop RG results \cite{hofstetter,beloborodov} from imaginary time formulation. While the thermodynamics of the ring type problem has been much studied, including extensive Monte Carlo studies \cite{herrero,lukyanov2} of $M^*$, no sign of a finite coupling fixed point has been detected. Our method evaluates the response to a strictly DC electric field $E$, equivalent to a magnetic flux through the ring that increases linearly with time, hence a non-equilibrium response. We claim that thermodynamic quantities like $M^*$, that are flux sensitive, decouple from the response to $E$, a response that averages over flux values.

In terms of the SEB, our results extend the previous analysis \cite{mora} to the case of many channels $N_c$ \cite{devoret}. We note that for $N_c > 1$ the relaxation resistance for noninteracting electrons becomes $h/(2N_ce^2)$ \cite{buttiker}.
We find that for strong coupling, $\eta/\hbar \gtrsim 1$ the relaxation resistance is quantized to $e^2/h$ up to an exponentially small correction $\sim \eexp{-\pi\eta/\hbar}$. For finite $\eta$, but still $\eta>\eta_c$ we find that a certain average of the relaxation resistance is quantized (see Eq. (\ref{e14}) below).

We proceed to reexamine the mapping of the box and ring problems. For the SEB one has the action
\beq{01}
S&=& \int_t\left\{\sum_{\alpha n}d^{\dagger}_{\alpha n}(i \hbar \partial_t-\epsilon_{\alpha})d_{\alpha n}
-E_c({\hat N}-N_0)^2\right\}\nonumber\\ &+& S_{lead}+S_{tun}
\eeq
where $d_{\alpha n}$ are dot electron operators, $n=1,..N_c$ labels the channels, ${\hat N}=\sum_{\alpha n}d^{\dagger}_{\alpha n}d_{\alpha n}$, $E_c=e^2/(2C_g)$ with $C_g$ is the geometric (bare) capacitance, $N_0$ is proportional to the gate voltage, $S_{lead}$ describes free electrons on the lead and $S_{tun}$ is the tunneling between the lead and the dot. We introduce an auxiliary variable $\theta_t$ with an action $E_c\int_t[{\hat N}-N_0- \hbar {\dot\theta}/2E_c]^2$ and rewrite the total action as
\beq{02}
S&=& \int_t\left\{\sum_{\alpha n}d^{\dagger}_{\alpha n}(i \hbar \partial_t-\epsilon_{\alpha}- \hbar {\dot\theta}_t)d_{\alpha n}
+\frac{\hbar^2 {\dot\theta}_t^2}{4E_c}+N_0 \hbar {\dot\theta}_t\right\}\nonumber\\ &+& S_{lead}+S_{tun}\,.
\eeq
 In terms of fermion operators ${\tilde d}_{\alpha n}=\eexp{i\theta(t)}d_{\alpha n}$, integrating out these fermions and expanding in $S_{tun}$ yields the well known effective action for the SEB \cite{AES,schon,golub,hofstetter,herrero,bulgadayev,lukyanov1,lukyanov2,burmistrov1,burmistrov2}. Eq. (\ref{e02}) shows that the equivalent particle on a ring has a mass $M=\hbar^2/(2E_c)$ (the radius of the ring is chosen as $=1$) and there is a flux (in unit of the flux quantum) $\phi_x=-N_0$ through the ring . The tunneling amplitudes squared, weighted by the number $N_c$ of channels, become the dissipation parameter $\eta$ of the particle. The mapping becomes exact in the large $N_c$ limit at fixed $\eta$ and for small mean level spacing \cite{beloborodov2} $\Delta\ll E_c$, a situation that can be realized \cite{devoret}; the application of this mapping is therefore limited to the temperature range $\Delta<T\ll E_c$.
Furthermore, by shifting $\hbar {\dot\theta}_t\rightarrow \hbar {\dot\theta}_t+2E_c({\hat N}_t-N_0)$ we obtain $\hbar \langle{\dot\theta}_t\rangle=2E_c[\langle{\hat N}\rangle_{N_0}-N_0]$ and also a relation between response functions
\beq{03}
\hbar^2 {\tilde K}_{t,t'}&=&-2E_c \hbar \delta(t-t')+4E_c^2K_{t,t'}
\eeq
where ${\tilde K}_{t,t'}=+i\theta(t-t')\langle [{\dot \theta}_t,{\dot\theta}_{t'}]\rangle$ is the response for the ring while $K_{t,t'}=+i\theta(t-t')\langle [{\hat N}_t,{\hat N}_{t'}]\rangle$ is for the SEB. The
$-2E_c \hbar \delta(t-t')$ term in (\ref{e03}) is essential, e.g. without tunneling the charge fluctuations are frozen, $K_{t,t'}=0$, while the corresponding particle is free with the correlation $-2E_c \hbar \delta(t-t')$.

The SEB response is parameterized as \cite{mora} $\frac{e^2}{\hbar}K(\omega)=C_0(1+i\omega C_0R_q)$ where $C_0$ is the effective DC capacitance and $R_q$ is the celebrated relaxation resistance \cite{buttiker}. The corresponding ${\tilde K}_{t,t'}$ is the response to a change in the external flux and is parameterized as
\beq{04}
{\tilde K}(\omega)=-K_0(\phi_x)+i\omega K_1(\phi_x) +O(\omega^2)
 \eeq
and the persistent current from a time independent flux is $\langle {\dot\theta}\rangle=\int_0^{\phi_x}K_0(\phi_x')d\phi_x'$. The continuation to imaginary time identifies the curvature of the free energy \cite{guinea,golubev1,kagalovsky,bh}, or an effective mass, as $\frac{1}{\hbar}\frac{\partial^2 F}{\partial \phi_x^2}=\hbar/M^*(\phi_x)=K_0(\phi_x)$; e.g. without tunneling $M^*=M$ while for large $\eta$ the effective mass $M^*\sim\eexp{\pi\eta/\hbar}$ is exponentially large.

Consider now the system in presence of a (classical) electric field $E$, of Hamiltonian
$\delta{\cal H}_{ring}=- (E + \delta E(t)) \theta$
and define the linear response $\delta \langle\theta_t\rangle_{E}=\int_{t'} {\cal R}_{t,t'} \delta E(t')$ to a small perturbation $\delta E$. This response is studied below for a DC field. In general its low frequency form is  (see (\ref{e10}) below) ${\cal R}(\omega)=\frac{-1}{i\omega\eta_R(E)}$ which defines $\eta_R(E)$ as a renormalized dissipation parameter.
 Since $E= \hbar {\dot\phi}_x$ we expect $\hbar\omega^2{\cal R}(\omega)={\tilde K}(\omega)$, hence the $K_0$ term in Eq. (\ref{e04}) is not reproduced. To resolve this discrepancy we note that an additional constant flux $\phi_x$ in the total flux $\phi_x+ E t/\hbar$ can be eliminated by redefining the origin of the time $t$, therefore the persistent current part should be eliminated. More precisely, define $\hbar \phi_x(t)=Et$; the 1st term in (\ref{e04}) $K_0(\phi_x)=K_0(Et/\hbar)$ becomes a periodic function, i.e. an AC response at $\omega_E=
 2\pi E/\hbar$. For a DC response at finite $E$ this persistent current response averages to zero, i.e. $\int_0^1K_0(\phi_x)d\phi_x=0$. The same reasoning applies to a $\phi_x$ average on $K_1(\phi_x)$.
 Hence the DC response to a DC field is given by
 \beq{05}
 \lim_{E \to 0} \lim_{\omega \to 0}\frac{{\tilde K}(\omega)}{\omega}=i\int_0^1K_1(\phi_x)d\phi_x\,.
 \eeq
 Therefore $\hbar/\eta_R=\int_0^1 K_1(\phi_x)d\phi_x$ where we denote $\eta_R\equiv \eta_R(E\rightarrow 0)$.
 The order of limits in (\ref{e05}) signifies that $\eta_R$ is essentially a non-equilibrium response.
The physical picture is that in a DC field the particle rotates around the ring and produces two types of currents. First is the persistent current that oscillates in time as $\phi_x$ increases and is therefore time averaged to zero; this current is non-dissipative. Second, there is a genuine DC response from the $i\omega K_1$ term, which is dissipative.

 In terms of the SEB response, using Eq. (\ref{e03}), we obtain the following mapping of ring and box parameters as functions of flux $\phi_x$ and $N_0$:
 \beq{06}
 \frac{M}{M^*(\phi_x)}&=&1-\frac{C_0(N_0)}{C_g}\nonumber\\
 \frac{\hbar}{\eta_R}&=&\frac{e^2}{\hbar}\int_0^1\frac{C_0^2(N_0)}{C_g^2}R_q(N_0)dN_0
 \eeq
 and we note also that $\int_0^1C_0(N_0)dN_0=C_g$.

 At this stage we can already propose an interesting experiment for the SEB. By analogy with $E= \hbar {\dot\phi}_x$ in the ring, we propose measuring the response to a gate voltage that is linear in time $N_0\sim t$. This leads to a DC current into the Coulomb box whose dissipation is the average in Eq. (\ref{e06}). This average is predicted to be quantized, at least for $\eta>\eta_c$, as discussed below.

We proceed now to study the ring problem.
To  derive the Keldysh action, we start from the well known action of a particle in a CL environment \cite{CL} in 2-dimensions with a position vector ${\bf x}^{\pm}$, where $\pm$ correspond to the upper and lower Keldysh contour,
\beq{07} S_K=i\int_{t,t'} {\hat {\bf x}}_t R^{-1}_{t,t'}{\bf x}_{t'} +\half
\int_{t,t'}{\hat {\bf x}}_tB_{t,t'}{\hat {\bf x}}_{t'}
\eeq
and ${\bf x}_t=\half({\bf x}_t^++{\bf x}_t^-)$ and ${\hat{\bf x}}_t=({\bf x}_t^+-{\bf x}_t^-)/\hbar$.
The simplest response function $R(\omega)$, in Fourier transform, and the noise function $B(\omega)$, at zero temperature, are \cite{supp}
\beq{08}
R(\omega)=\frac{-1}{M\omega^2+i\eta\omega},\qquad B(\omega)=\hbar\eta|\omega|
\eeq
 This quadratic problem corresponds to a particle of mass $M$ and a friction $\eta$ within a Langevin equation $M\ddot{{\bf x}}+\eta \dot{{\bf x}}={\bm \xi}_t$; each component of ${\bm \xi}_t=(\xi^x_t,\xi^y_t)$ is random with correlations $B(\omega)$.

 We project now the position on a ring, i.e. ${\bf x}^{\pm}_t=(\cos\theta^{\pm}_t,\sin\theta^{\pm}_t)$, and rewrite the action in terms of classical and quantum angle variables $\theta_t=\half(\theta^+_t+\theta^-_t)$ and ${\hat\theta}_t=(\theta^+_t-\theta^-_t)/\hbar$ :
\beq{09}
S_K&=&S_0+S_{int}+S_c
\nonumber\\
S_0&=&i\int_{t,t'} {\hat\theta}_t R^{-1}_{tt'}\delta\theta_{t'} = i\int_{t,t'} {\hat\theta}_t R^{-1}_{tt'} \theta_{t'} - i E \int_t {\hat\theta}_t \nonumber\\
S_{int}&=&\frac{2}{\hbar^2}\int_{t,t'}B_{t,t'}\sin(\frac{\hbar}{2}{\hat
\theta_t})\sin(\frac{\hbar}{2}{\hat \theta}_{t'})\cos(\theta_{t'}-\theta_t)\nonumber\\
S_c
&=&\frac{i\eta}{\hbar}\int_t[\sin(\hbar{\hat\theta}_t){\dot\theta}_{t^-}
-\hbar{\hat\theta}_t{\dot\theta}_{t^-}]
\eeq
where the last term assumes the form (\ref{e08}) and $t^-$ is infinitesimal below $t$. A Gaussian term $S_0$  has been singled out so that a perturbation scheme in powers of $S_{int}, S_c$ can be defined.
We have added an external electric field $E$, hence the particle acquires a velocity $v=\langle {\dot\theta}_t\rangle$ as a function of $E$.
To perform a perturbation theory it is convenient to introduce the bare velocity $v_0=E/\eta$ and to define $\theta_t=\delta\theta_t+v_0t$.
 The derivative of the $v(E)$ characteristics is easily shown to be related to
$\eta_R(E)$ via:
\beq{10}
\frac{dv}{dE}&=&i\bra \int_{t'}{\dot\theta}_t{\hat\theta}_{t'}\ket=
\lim_{t-t'\rightarrow\infty}{\cal R}_{t,t'} \equiv \frac{1}{\eta_R(E)}
\eeq
where ${\cal R}_{t,t'}=i\bra \theta_t{\hat\theta}_{t'}\ket$ is the full response function defined above. We note that the form (\ref{e09}) for $S_K$ has been derived also for the SEB \cite{AES,schon,golub,burmistrov2,burmistrov3}.

The semiclassical limit of (\ref{e09}), which corresponds to small $\hbar/\eta$, is obtained by linearizing the sine terms, and is
equivalent to a Langevin equation (also obtained for the SET \cite{golubev2})
\beq{11}
 M\ddot{\theta}_t +\eta \dot{\theta}_t=\xi^x_t\cos\theta+\xi^y_t\sin\theta_t  +E
 \eeq
which is in fact the 2D Langevin equation projected on the tangent to the ring.

We perform a perturbative expansion of the action with respect to $S_{int}, S_c$ to compute $\eta_R(E)$.
The perturbative expansion of $\eta_R(E)$ exhibits logarithmic divergences when $E \to 0$. The velocity $v_0$ thus provides a natural low frequency cutoff for this divergences, and the mass provides a high frequency cutoff at $\omega_c=\eta/M$. The expansion terms can be classified as n-loops by looking at the small $\hbar/\eta$ power of each term which is of order $R^{2n-1}B^n/\eta^2\sim \hbar^n/\eta^{n+1}$. However we find, due to the periodicity of the action in the angle variables, that the $R^{2n-1}$ factors in front of the logarithmic terms become periodic functions:
The result up to 2-loops and $O(v_0)$ is \cite{supp}:
\beq{12}
\frac{1}{\eta_R(E)}&=&\frac{1}{\eta}-\frac{2}{\pi\eta}\sin(\frac{\hbar}{2\eta})\ln[v_0/\omega_c']\\
&+&
\frac{4}{\pi^2\hbar}\sin^2(\frac{\hbar}{2\eta})\sin(\frac{\hbar}{\eta})
\{\ln^2[v_0/\omega_c']+b_0\ln[v_0/\omega'_c]\}\nonumber
\eeq
where $b_0=O(1)$ may weakly depend on $\eta$ and
$\omega_c'/\omega_c=1+O(1/\eta^2)$.
In the limit of large $\eta$ one
can reexpress (\ref{e12}) in terms of the small parameter $\gamma=\frac{\hbar}{\pi \eta}$ and $\gamma_R=\frac{\hbar}{\pi \eta_R(E)}$
and obtain the 2-loop $\beta$ function as $- E \partial_E \gamma_R = \gamma_R^2 - b_0 \gamma_R^3+ O(\gamma_R^4)$ which has the same form as from the semi-classical equation.
We show in Fig. 1 our numerical solution for Eq. (\ref{e11}) with a reasonable fit to the 2-loop form with
$b_0=0$. When $1/v_0$ approaches the simulation time span the numerics, and the plateau observed at low $E$,
become unreliable. We note that a similar 2-loop result was found in equilibrium \cite{hofstetter,beloborodov} with $b_0=-1$. The full quantum theory (\ref{e09}) including its non-equilibrium limit (\ref{e05}) differs from these descriptions \cite{hofstetter,beloborodov, burmistrov3}.

\begin{figure}[h]
\begin{center}
\includegraphics[scale=0.4]{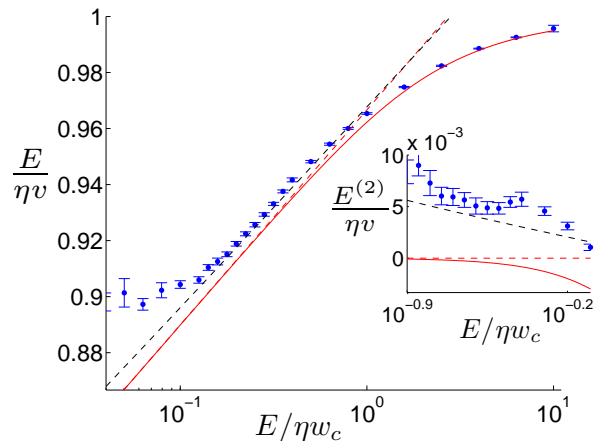}
\end{center}
\caption{Velocity-field relation for Eq. (\ref{e11}) with $\eta=30\hbar/\pi$. The circles are numerical data, the full line is a 1st order
perturbation in $1/\eta$, the dashed lower (red) line is its logarithmic expansion for large $\ln v_0/\omega_c$ ($v_0=E/\eta$ being the bare velocity) and the dashed upper (black) line includes the 2nd order logarithmic term, corresponding to Eq. (\ref{e12}) for $\hbar\rightarrow 0$ and $b_0=0$. The 2nd order terms are also shown in the inset after the 1st order is subtracted, i.e. $\frac{E^{(2)}}{\eta v}= \frac{E}{\eta v}-1-\frac{\hbar}{\pi\eta}(\ln \frac{v_0}{\omega_c}-1)$.}
\end{figure}

 We consider now the quantum theory, beyond large $\eta$. We note that in (\ref{e12}) $g=\frac{2}{\pi}\sin(\frac{\hbar}{2\eta})$ acts as an unexpected small parameter for the expansion, since all
divergences vanish when $g=0$. It raises the interesting possibility that $g=0$ be viewed as a RG fixed point.
For that we need to find a renormalized coupling which obeys multiplicative RG, the simplest choice being
$g_R=\frac{2}{\pi}\sin(\frac{\hbar}{2\eta_R(E)})$. The question is then whether the $\beta$-function $\beta=- E \partial_E g_R$ can be written only in terms of $g_R$. Although the non-periodic $1/\eta$ factor in (\ref{e11}) appears at first problematic, we propose that resummation from higher loops, which allows for higher order terms $O(\frac{1}{\eta^4})$ changes the 1-loop term in (\ref{e12}) by $\frac{\hbar}{2\eta}\rightarrow\sin(\frac{\hbar}{2\eta})$, so that by taking a sine of both sides it yields to order $g^3$
\beq{13}
g_R=g\pm g^2\ln(v_0/\omega'_c)+g^3[\ln^2(v_0/\omega'_c)+b_0\ln(v_0/\omega'_c)] \nonumber\\
\eeq
where $\pm$ refers to $g=0$ with $\cos(\frac{\hbar}{2\eta})=\pm 1$, leading to $\beta(g_R)=\mp g_R^2 - b_0 g_R^3+ O(g_R^4)$.

To further motivate this proposal we consider the response ${\bar R}_{t,t'}=i\frac{2}{\hbar}\bra \theta_t\sin(\frac{\hbar}{2}{\hat\theta}_{t'})\ket$. Physically, $\eexp{\pm i \frac{\hbar}{2}{\hat\theta}_{t'}}$  corresponds to an electric field pulse
$\delta E(t)=\pm\frac{\hbar}{2}\delta (t-t')$ or equivalently a rapid change of flux by $\pm\half$, therefore
${\bar R}_{t,t'}$ corresponds to the difference in response to these two flux pulses. For ${\bar R}_{t,t'}$  the 1-loop term is fully periodic with $\frac{\hbar}{2\eta}\rightarrow\sin(\frac{\hbar}{2\eta})$ in Eq. (\ref{e12}). We note that there are many other operators that have vanishing perturbations at $g=0$ to 2nd order in $S_{int},S_c$, e.g. the dissipation term in Eq. (\ref{e09}) $\bra \theta_t\sin(\hbar{\hat\theta}_{t'})\ket$, or the response to an AC field with frequency $v$
$\langle \theta_t\cos\delta\theta_{t'}\sin\frac{\hbar}{2}{\hat\theta}_{t'}\rangle$.
  We propose then that $g=0$ are exact zeroes of the perturbation expansion and requiring an RG structure leads then to the result (\ref{e13}).

Eq. (\ref{e12}) yields fixed points at $\frac{\hbar}{2\eta_n}=n\pi$ with $n=1,2,3,...$ that are attractive at $\eta>\eta_n$ and repulsive at $\eta<\eta_n$, i.e. the flow of $\eta\neq\eta_n$ is always to smaller $\eta$. At these fixed points a Gaussian evaluation yields the correlation $\langle\cos\theta_t\cos\theta_0\rangle\sim t^{-2n}$. We recall now a theorem for the lattice model \cite{spohn} where the equilibrium action with mass related cutoff is replaced by an action on a lattice resulting in an XY model with long range interactions. The theorem states \cite{spohn} that $\langle\cos\theta_t\cos\theta_0\rangle\sim 1/t^2$; this result was also derived in first order in $\eta$ \cite{bh}. The range $\eta>\eta_1$ has an RG flow to $\eta_1$ and is therefore consistent with the theorem. The hypothesis of Gaussian fixed points corresponding to $n \geq 2$ is inconsistent with the theorem, i.e. $\langle\cos\theta_t\cos\theta_0\rangle$ becomes a relevant operator at the $n\leq 2$ points rendering them unstable. For $\eta < \eta_1$ the system may have
non-gaussian fixed points or a line of fixed points as hinted by the small $\eta$ perturbation \cite{bh}.
Note that in the SEB problem $\cos\theta_t$ corresponds to a lead-dot voltage and its correlations determine the SET conductance \cite{AES,schon,burmistrov1}, while in the ring problem it corresponds to fluctuations in the circular asymmetry.

The special value $\eta_R=\hbar/(2 \pi)$ has a topological interpretation as a Thouless charge pump \cite{thouless}. Consider a slow change of $\phi_x$ by one unit with $\hbar{\dot \phi}_x=\eta_R\langle{\dot\theta}\rangle$. For this
special value $\eta_R=\hbar/(2 \pi)$ the total change in the position of the particle
$\int_t \langle{\dot\theta}\rangle dt=2\pi$, i.e. the particle comes back to the same position on the ring and a unit charge has been transported. Such quantization has been shown for cases where the spectrum has a gap \cite{thouless}, though quantized charge transport was shown also in cases without a gap \cite{aleiner,sharma}. The quantized $\eta_R$ also results from arguing that there should be a unique frequency $\omega_E = v$ as $E \to 0$, as suggested by linear response.

We conclude that for $\eta>\eta_1\equiv\eta_R$ the SEB satisfies the quantization
\beq{14}
\int_0^1\frac{C_0^2(N_0)}{C_g^2}R_q(N_0)dN_0=\frac{h}{e^2}\,.
\eeq
In particular, when $\eta/\hbar \gtrsim 1$ we have from the known $M^*/M\sim\eexp{\pi\eta/\hbar}$ \cite{guinea,golubev1,kagalovsky,bh} and from Eq. (\ref{e06}) that $C_0/C_g=1+O(\eexp{-\pi\eta/\hbar})$. We expect $R_q$ to be independent of $N_0$ at large $\eta$, hence
\beq{15}
R_q=\frac{h}{e^2}[1+O(\eexp{-\pi\eta/\hbar})]
\eeq
similar to the $N_c=1$ case \cite{mora}.

The conductance for the ring can be defined by the voltage around the ring  $2\pi E/e$ and the current  $e\langle\dot\theta\rangle/2\pi$, hence we expect the conductance for $\eta>\eta_R$ to be:
 \beq{16}
 G_{ring}=\frac{e^2}{4\pi^2\eta_R} =\frac{e^2}{h}\,.
 \eeq

 Finally, we reconsider the conditions for our proposed box experiment. The field $E$ should be sufficiently small so that $g_R$ is sufficiently near the fixed point. For an initial $g\approx 1$ integration of
 $\partial g_R/\partial \ln E=g_R^2$ yields $g_R=1/\ln (\hbar \omega_c/E) \ll g$. E.g. for $g_R\lesssim 0.1$ and a typical $\hbar \omega_c\approx 1$meV one needs $E/\hbar\lesssim 10^8$Hz. $E/\hbar$ has frequency units, corresponding to $10^8$ electrons/sec flowing into the box. We propose measuring the charge fluctuations (noise) $S_{Q}(\omega)=e^2\langle {\hat N}_t{\hat N}_{t'}\rangle_{\omega}$
 at a frequency, temperature and level spacings $\Delta$ such that $\Delta<\omega, T \ll 10^8$Hz, to yield the DC response (\ref{e05},\ref{e14})). We predict then that the noise $S_{Q}(\omega)(\frac{2E_c}{e\hbar})^2\frac{1}{\omega}=
 \frac{\hbar}{\eta_R}=2\pi$ is quantized.

 Acknowledgements: We thank G. F\`{e}ve, A. Golub, D. Goldhaber-Gordon, S. L. Lukyanov, Y. Meir, C. Mora, B. Pla\c{c}ais and G. Zar\'{a}nd for stimulating discussions. BH acknowledges kind hospitality and financial support from LPTENS and PLD from Ben Gurion University. This research was supported by THE ISRAEL SCIENCE FOUNDATION (grant No. 1078/07) and by the ANR grant 09-BLAN-0097-01/2.

\end{document}